\documentclass[12pt]{article}
\usepackage{amssymb,a4,epsf}
\usepackage{amsmath}

\begin{document}

\centerline{\Large\bf Colourings of Planar Quasicrystals}\vspace{5mm}

\centerline{\large\sc Michael Baake$^{(1)}$, 
Uwe Grimm$^{(2)}$, and Max Scheffer$^{(3)}$}\vspace{5mm}

\begin{small}
\centerline{%
\begin{minipage}{0.7\textwidth}
(1) Institut f\"{u}r Mathematik, Universit\"{a}t Greifswald,\\
\hphantom{(1)} Jahnstr.~15a, 17487 Greifswald, Germany\\
(2) Applied Mathematics Department, The Open University,\\ 
\hphantom{(2)} Walton Hall, Milton Keynes MK7 6AA, UK\\
(3) Institut f\"{u}r Physik, Technische Universit\"{a}t,\\ 
\hphantom{(3)} 09107 Chemnitz, Germany
\end{minipage}}
\end{small}\vspace{5mm}

\begin{abstract}
\noindent
The investigation of colour symmetries for periodic and aperiodic
systems consists of two steps. The first concerns the computation of
the possible numbers of colours and is mainly combinatorial in
nature. The second is algebraic and determines the actual colour
symmetry groups. Continuing previous work, we present the results of
the combinatorial part for planar patterns with $n$-fold symmetry,
where $n=7,9,15,16,20,24$. This completes the cases with $\phi(n)\le
8$, where $\phi$ is Euler's totient function.
\end{abstract}\vspace{5mm}

Colour symmetries of crystals and, more recently, of quasicrystals
continue to attract a lot of attention, simply because so little is
known about their classification, see \cite{L1} for a recent review.
A first step in this analysis consists in answering the question how
many different colourings of an infinite point set exist which are
compatible with its underlying symmetry. More precisely, one has to
determine the possible numbers of colours, and to count the
corresponding possibilities to distribute the colours over the point
set (up to permutations).

In this generality, the problem has not even been solved for simple
lattices. One common restriction is to demand that one colour occupies
a subset which is of the same Bravais type as the original set, while
the other colours encode the cosets. In this situation, to which we will
also restrict ourselves, several results are known and can be given in
closed form \cite{B,BHLS,BM,L1,L2}. Of particular interest are planar
cases because, on the one hand, they show up in quasicrystalline
$T$-phases, and, on the other hand, they are linked to the rather
interesting classification of planar Bravais classes with $n$-fold
symmetry \cite{MRW}. They are unique for the following 29 choices of
$n$:
\begin{eqnarray*}
n&=&3,4,5,7,8,9,11,12,13,15,16,17,19,20,21,24,\\
& & 25,27,28,32,33,35,36,40,44,45,48,60,84.
\end{eqnarray*}
The canonical representatives are the sets of cyclotomic integers
${\cal M}_{n}=\mathbb{Z}[\xi_{n}]$ where $\xi_{n}=\exp(2\pi i/n)$ with
$n$ from our list.  Except $n=1$ (where ${\cal M}_{1}=\mathbb{Z}$ is
one-dimensional), these are all cases where $\mathbb{Z}[\xi_{n}]$ is a
principal ideal domain and thus has class number one, see \cite{W,BM}
for details. If $n$ is odd, we have ${\cal M}_{n} = {\cal M}_{2n}$ and
${\cal M}_{n}$ thus has $2n$-fold symmetry. To avoid duplication of
results, values $n\equiv 2$ (mod $4$) do not appear in the above
list. There are systematic mathematical reasons to prefer this
traditional convention to the notation of \cite{L1,MRW}.

It is this very connection to algebraic number theory which gives the
Bravais classification \cite{MRW}, and also allows for a solution of
the combinatorial part of the colouring problem by means of Dirichlet
series generating functions. The latter is explained in \cite{B},
where the solutions for all $n$ with $\phi(n)\le 4$ and for $n=7$ are
given explicitly. Here, $\phi(n)$ is Euler's totient function which is
the number of integers $k$, $1\le k\le n$, which are coprime with $n$.
Note that $\phi(n)=2$ are the crystallographic cases $n=3$ (triangular
lattice) and $n=4$ (square lattice), while $\phi(n)=4$ means
$n\in\{5,8,12\}$ which are the standard symmetries of planar
quasicrystals. Again, $n=10$ is covered implicitly, as explained above.

The methods emerging from this connection allow the full treatment of
all 29 cases listed above, and this will be spelled out in detail in a
forthcoming publication. Here, we present the results for $\phi(n)=6$
(i.e., $n\in\{7,9\}$) and for $\phi(n)=8$ (i.e.,
$n\in\{15,16,20,24\}$), thus completing the cases with $\phi(n)\le 8$,
where partial results had been given in \cite{B,L1}.

Let us consider ${\cal M}_n$ with a fixed $n$ from our list.  Then,
${\cal M}_n$ is an Abelian group, and also a $\mathbb{Z}$-module of
rank $\phi(n)$. We view it as a subset of the Euclidean plane, and
hence as a geometric object. Our combinatorial problem is now to
determine the values of the multiplicative arithmetic function
$a^{}_{n}(k)$ which counts the possibilities to colour an $n$-fold
symmetric submodule of ${\cal M}_n$ and its cosets with $k$ different
colours, see \cite{B,L1} for details.  These submodules are then
necessarily principal ideals of ${\cal M}_n = \mathbb{Z}[\xi_n]$,
i.e., sets of the form $g {\cal M}_n$ with $g\in {\cal M}_n$.

On the other hand, for the $n$ from the above list, {\em all}\/ ideals
of ${\cal M}_n$ are principal, so our combinatorial problem amounts to
counting all (non-zero) ideals, which is achieved by the Dedekind
zeta-function of the cyclotomic field $\mathbb{Q}(\xi_n)$. The number
of colours then corresponds to the norm of the ideal \cite{BM}, and we 
obtain the following Dirichlet series generating function
\begin{equation}
 \zeta^{}_{{\cal M}_n} (s) \; := \;
 \sum_{k=1}^{\infty} \frac{a^{}_{n}(k)}{k^s} \; = \;
 \zeta^{}_{\mathbb{Q}(\xi_n)} (s) \; = \;
 \prod_{p \; {\rm prime}} E(p^{-s})\, .
 \label{euler}
\end{equation}
The last expression is called the Euler product expansion of the Dirichlet
series. Each Euler factor is of the form
\begin{equation}
 E(p^{-s}) \; = \; \frac{1}{(1-p^{-\ell s})^m} \; = \;
 \sum_{k=0}^{\infty} \binom{k+m-1}{m-1}
 \frac{1}{(p^{s})^{k\ell}}
 \label{factor}
\end{equation}
from which one deduces the value of $a^{}_{n}(p^r)$ for $r\ge 0$.
The indices $\ell,m$ depend on the prime $p$, and on the choice of
$n$.

If $p\equiv k$ (mod $n$) with $k$ and $n$ coprime, then $\ell\cdot m =
\phi(n)$. Such primes are listed as $p_{k}^{\ell}$ in Table
\ref{tab:basic}.  In addition, there are finitely many primes $p$
which divide $n$, the so-called ramified primes. They are listed as
$p^{\ell}$ in Table \ref{tab:basic}, and here we have $m=1$ except for
two cases (where $m=2$). With this information, one can easily
calculate the possible numbers of colours and the generating functions
by inserting (\ref{factor}) into (\ref{euler}) and expanding the Euler
product.

\begin{table}
\begin{center}\renewcommand{\arraystretch}{1.3}
\begin{tabular}{|c|r|l|l|}
 \hline
 $\phi(n)$ & $n$ & special $p$ & \multicolumn{1}{c|}{general $p$} \\
 \hline
 2 &  3 & 3            & $p_{1}^{}$, $p_{2}^{2}$  \\
   &  4 & 2            & $p_{1}^{}$, $p_{3}^{2}$ \\
 \hline
 4 &  5 & 5            & $p_{1}^{}$, $p_{2}^{4}$, 
                         $p_{3}^{4}$, $p_{4}^{2}$\\
   &  8 & 2            & $p_{1}^{}$, $p_{3}^{2}$, 
                         $p_{5}^{2}$, $p_{7}^{2}$\\
   & 12 & $2^2$, $3^2$ & $p_{1}^{}$, $p_{5}^{2}$, 
                         $p_{7}^{2}$, $p_{11}^{2}$\\
 \hline
 6 &  7 & 7            & $p_{1}^{}$, $p_{2}^{3}$, 
                         $p_{3}^{6}$, $p_{4}^{3}$, 
                         $p_{5}^{6}$, $p_{6}^{2}$\\
   &  9 & 3            & $p_{1}^{}$, $p_{2}^{6}$, 
                         $p_{4}^{3}$, $p_{5}^{6}$, 
                         $p_{7}^{3}$, $p_{8}^{2}$\\
 \hline
 8 & 15 & $3^4$, $5^2$ & $p_{1}^{}$, $p_{2}^{4}$, 
                         $p_{4}^{2}$, $p_{7}^{4}$, 
                         $p_{8}^{4}$, $p_{11}^{2}$, 
                         $p_{13}^{4}$, $p_{14}^{2}$\\
   & 16 & 2            & $p_{1}^{}$, $p_{3}^{4}$, 
                         $p_{5}^{4}$, $p_{7}^{2}$, 
                         $p_{9}^{2}$, $p_{11}^{4}$, 
                         $p_{13}^{4}$, $p_{15}^{2}$\\
   & 20 & $2^4$, $5\, [m\! =\! 2]$   & 
                         $p_{1}^{}$, $p_{3}^{4}$, 
                         $p_{7}^{4}$, $p_{9}^{2}$, 
                         $p_{11}^{2}$, $p_{13}^{4}$, 
                         $p_{17}^{4}$, $p_{19}^{2}$\\
   & 24 & $2^2$, $3^2\, [m\!=\!2]$  & 
                         $p_{1}^{}$, $p_{5}^{2}$, 
                         $p_{7}^{2}$, $p_{11}^{2}$, 
                         $p_{13}^{2}$, $p_{17}^{2}$,  
                         $p_{19}^{2}$, $p_{23}^{2}$\\
 \hline
\end{tabular}\renewcommand{\arraystretch}{1}
\caption{Basic indices for the numbers of colours. Composite
 numbers (i.e., possible numbers of colours) 
 are all products of basic indices.\label{tab:basic}}
\end{center}
\end{table}

The result for $n=7$ was already given in equation (11) of reference 
\cite{B}. The Euler product is correct, but the explicit first few terms 
contain one misprint, which we correct here
\[
 \zeta_{{\cal M}_{7}}^{}(s)\; =\; 
 \mbox{\small $  
 1 + \frac{1}{7^s} + \frac{2}{8^s} + \frac{6}{29^s} +
 \frac{6}{43^s} + \frac{1}{49^s} + \frac{2}{56^s} + \frac{3}{64^s} +
 \frac{6}{71^s} + \frac{6}{113^s} + \frac{6}{127^s} + \frac{3}{169^s} +
 \ldots $}
\]
The case $n=9$, in full detail, reads
\begin{eqnarray*}
 \zeta_{{\cal M}_9}^{}(s) & = & \frac{1}{1-3^{-s}} 
 \prod_{p\equiv 1\bmod 9} \frac{1}{(1-p^{-s})^{6}}
 \prod_{p\equiv 2,5\bmod 9} \frac{1}{1-p^{-6s}}\\
 & &   \cdot \prod_{p\equiv 4,7\bmod 9} \frac{1}{(1-p^{-3s})^{2}}
      \prod_{p\equiv 8\bmod 9} \frac{1}{(1-p^{-2s})^{3}}\\
 & = & \mbox{\small $ 1 + \frac{1}{3^s} + \frac{1}{9^s} + \frac{6}{19^s} + 
       \frac{1}{27^s} + \frac{6}{37^s} + \frac{6}{57^s} + 
       \frac{1}{64^s} + \frac{6}{73^s} + \frac{1}{81^s} + 
       \frac{6}{109^s} + \frac{6}{111^s} + \ldots $}
\end{eqnarray*}
The remaining cases can be calculated in the same way, using the data
of Table~\ref{tab:basic}. Here, we just spell out the first few
terms of the Dirichlet series for the four solutions of $\phi(n)=8$:
\begin{eqnarray*}
 \zeta_{{\cal M}_{15}}^{}(s) 
 & = & \mbox{\small $ 1 + \frac{2}{16^s} + \frac{1}{25^s} + 
       \frac{8}{31^s} + 
       \frac{8}{61^s} + \frac{1}{81^s} + \frac{4}{121^s} + 
       \frac{8}{151^s} + \frac{8}{181^s} + \frac{8}{211^s} + 
       \frac{8}{241^s} + \ldots $} \\
 \zeta_{{\cal M}_{16}}^{}(s) 
 & = & \mbox{\small $1 + \frac{1}{2^s} + \frac{1}{4^s} + \frac{1}{8^s} + 
       \frac{1}{16^s} + \frac{8}{17^s} + \frac{1}{32^s} + 
       \frac{8}{34^s} + \frac{4}{49^s} + \frac{1}{64^s} +
       \frac{8}{68^s} + \frac{2}{81^s} + \ldots $}
       \nonumber\\
 \zeta_{{\cal M}_{20}}^{}(s) 
 & = & \mbox{\small $ 1 + \frac{2}{5^s} + \frac{1}{16^s} + 
       \frac{3}{25^s} + 
       \frac{8}{41^s} + \frac{8}{61^s} + \frac{2}{80^s} + 
       \frac{2}{81^s} + \frac{8}{101^s} + \frac{4}{121^s} +
       \frac{4}{125^s} + \ldots $}  \\
 \zeta_{{\cal M}_{24}}^{}(s) 
 & = & \mbox{\small $ 1 + \frac{1}{4^s} + \frac{2}{9^s} + 
       \frac{1}{16^s} + 
       \frac{4}{25^s} + \frac{2}{36^s} + \frac{4}{49^s} + 
       \frac{1}{64^s} + \frac{8}{73^s} + \frac{3}{81^s} +
       \frac{8}{97^s} + \frac{4}{100^s} +  \ldots $}
\end{eqnarray*}

\begin{figure}
\centerline{\epsfxsize 15cm\epsffile{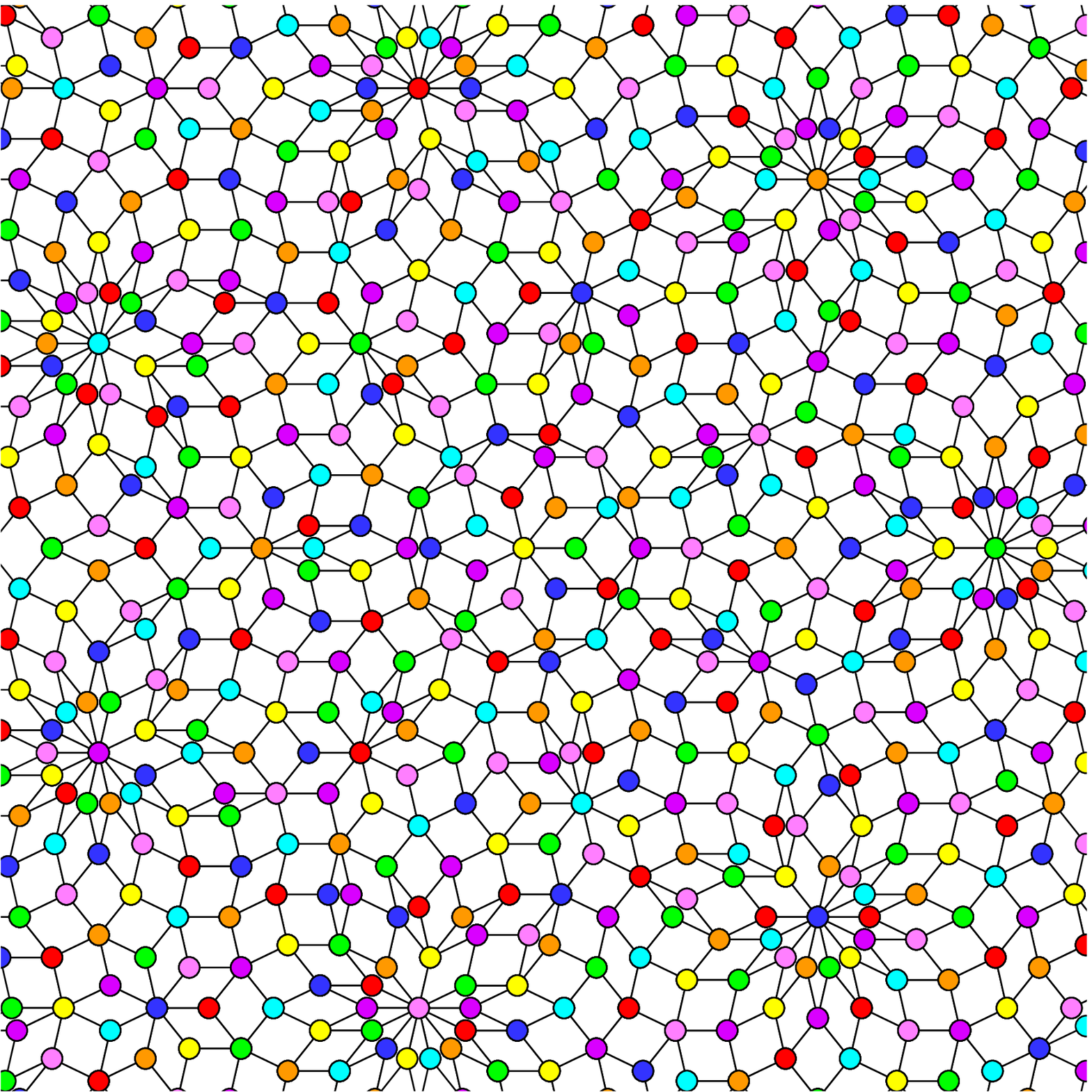}}\vspace{8mm}
\centerline{\epsfxsize 15cm\epsffile{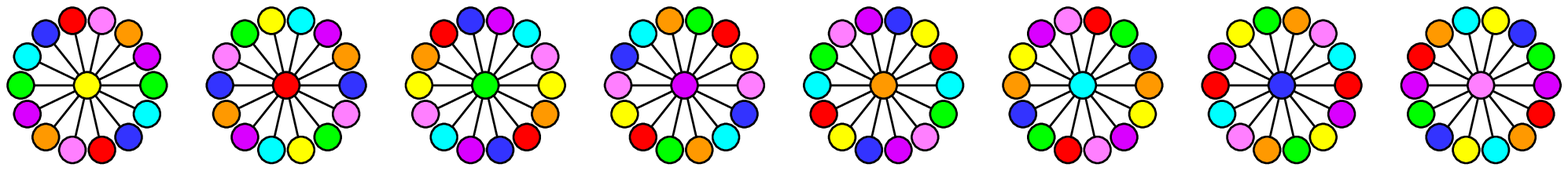}}\vspace{5mm}
\caption{Vertex colouring of a sevenfold tiling with eight colours.
The eight colour stars encode the relations between the colours of 
neighbouring vertices.\label{fig1}}
\end{figure}

\begin{figure}
\centerline{\epsfxsize 15cm\epsffile{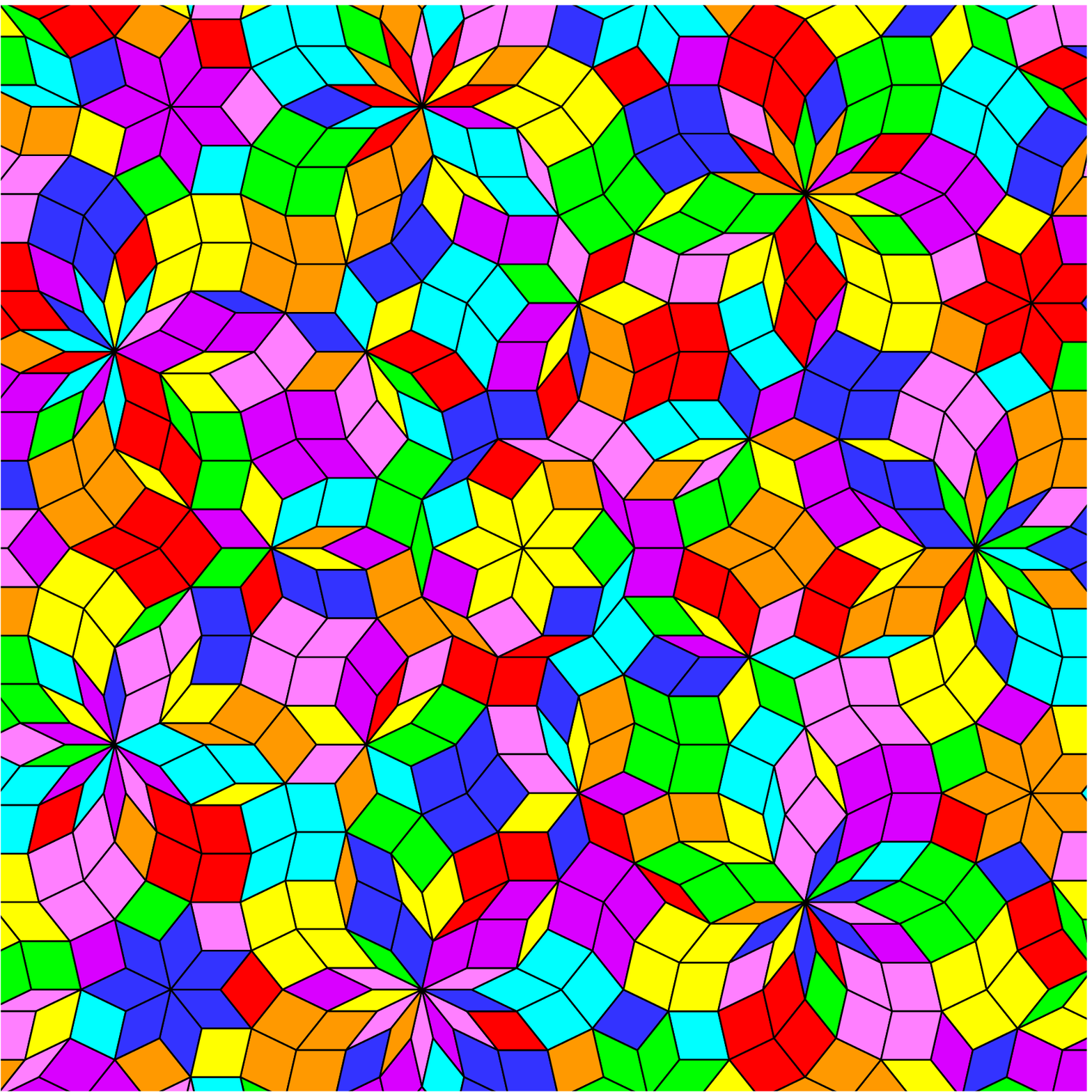}}\vspace{8mm}
\caption{Colouring of a sevenfold tiling with eight colours.\label{fig2}}
\end{figure}

To illustrate our findings, we present a colouring of a sevenfold
rhombus tiling with eight colours. In Figure~\ref{fig1}, the vertices
of the tiling are coloured, which are a subset of the module ${\cal
M}_{7}$.  
This case corresponds to one of the two possible colouring with eight 
colours (up to permutations), the other inequivalent colouring being 
obtained by a reflection.
Alternatively, one can colour the tiles (which is some sort
of complementary problem), by locally assigning a point of ${\cal
M}_{7}$ to each tile and transferring the corresponding colour. An
example is shown in Figure~\ref{fig2}. 
\bigskip

The authors gratefully acknowledge fruitful discussions with
Reinhard L\"{u}ck and Robert V.\ Moody.

\clearpage

\clearpage

\begin{thebibliography}{99}\itemsep 0pt

\bibitem{B}
M.~Baake,
Combinatorial aspects of colour symmetries,
{\em J.\ Phys.\ A: Math.\ Gen.}\/ {\bf 30} (1997) 2687--2698.

\bibitem{BHLS}
M.~Baake, J.~Hermisson, R.~L\"{u}ck, and M.~Scheffer,
Colourings of quasicrystals, in: 
{\em Quasicrystals}, eds.\ S.~Takeuchi and T.~Fujiwara
(World Scientific, Singapore 1998), pp.~120--123.

\bibitem{BM}
M.~Baake and R.~V.~Moody,
Similarity submodules and semigroups, in:
{\em Quasicrystals and Discrete Geometry}, ed.\ J.~Patera,
Fields Institute Monographs, vol.~10 (AMS, Providence, RI 1998),
pp.~1--13.

\bibitem{L1}
R.~Lifshitz,
Theory of color symmetry for periodic and quasiperiodic crystals,
{\em Rev.\ Mod.\ Phys.}\/ {\bf 69} (1997) 1181--1218.

\bibitem{L2}
R.~Lifshitz,
Lattice color groups of quasicrystals, in: 
{\em Quasicrystals}, eds.\ S.~Takeuchi and T.~Fujiwara
(World Scientific, Singapore 1998), pp.~103--107.

\bibitem{MRW}
N.~D.~Mermin, D.~S.~Rokhsar, and D.~C.~Wright,
Beware of $46$-fold symmetry: The classification of two-dimensional
quasicrystallographic lattices,
{\em Phys.\ Rev.\ Lett.}\/ {\bf 58} (1987) 2099--2101.

\bibitem{W}
L.~C.~Washington, 
{\em Introduction to Cyclotomic Fields}, 2nd ed.\
(Springer, New York, 1997).

\end{thebibliography}
\end{document}